\documentclass[journal]{IEEEtran}
\IEEEoverridecommandlockouts
\usepackage{cite}
\usepackage{amsmath,amssymb,amsfonts}
\usepackage{mathrsfs}
\usepackage{algorithmic}
\usepackage{graphicx}
\usepackage{textcomp}
\usepackage{amssymb}
\usepackage{xcolor}
\usepackage{float}
\usepackage{caption}
\usepackage{subfigure}
\usepackage{algorithm}
\usepackage{diagbox}
\usepackage{epstopdf}
\usepackage{hyperref}
\allowdisplaybreaks[4]
\hypersetup{hypertex=true,
            colorlinks=true,
            linkcolor=blue,
            anchorcolor=blue,
            citecolor=blue}
\def\BibTeX{{\rm B\kern-.05em{\sc i\kern-.025em b}\kern-.08em
    T\kern-.1667em\lower.7ex\hbox{E}\kern-.125emX}}
\begin{document}

\title{Joint Frame Structure, Beamwidth, and Power Allocation for UAV-Aided Localization and Communication}


\author{Tianhao Liang, \IEEEmembership{Graduate Student Member,~IEEE}, Tingting Zhang, \IEEEmembership{Member,~IEEE},  Sheng Zhou, \IEEEmembership{Member,~IEEE}, Wentao Liu, Dong Li, \IEEEmembership{Senior Member,~IEEE}, Qinyu Zhang, \IEEEmembership{Senior Member,~IEEE}

\thanks{Part of this work will be presented in the IEEE WCNC 2024 \cite{LTHWCNC}

Tianhao Liang, Tingting Zhang and Qinyu Zhang are with School of Electronics and Information Engineering, Harbin Institute of Technology (Shenzhen), Shenzhen, China  (e-mail: liangth@stu.hit.edu.cn; zhangtt@hit.edu.cn; zqy@hit.edu.cn).

Sheng Zhou is with Department of Electronic Engineering, Tsinghua University, Beijing, China (e-mail: sheng.zhou@tsinghua.edu.cn).

Wentao Liu and Dong Li are with the School of Computer Science and Engineering, Macau University of Science and Technology, Macau, China (e-mail: 2202853eii30001@student.must.edu.cn;  dli@must.edu.mo).
 
}}
\maketitle

\begin{abstract}
In wireless sensors networks, integrating localization and communications techniques is crucial for efficient spectrum and hardware utilizations. 
In this paper, we present a novel framework of unmanned aerial vehicle (UAV)-aided localization and communication for ground node (GN), where the average spectral efficiency (SE) is used to reveal the intricate relationship among frame structure, channel estimation error, and localization accuracy. In particular, we first derive the lower bounds for channel estimation error and the three dimensional location prediction error. Leveraging these comprehensive analysis, we formulate a problem to maximize the average SE in UAV-GN communication, where the frame structure, beamwidth and power allocation are jointly optimized. Subsequently, we propose an efficient iterative algorithm to address this non-convex problem with closed-form expressions for beamwidth and power allocation.  Numerical results demonstrate that the performance of our proposed method can approach the upper bound with much lower complexity,  and achieve over 70\% performance gain compared to non-localization benchmarks. Additionally, the analysis highlights the dominated impacts from the Doppler effect over noise on the average SE.
\end{abstract}

\section{Introduction}
\subsection{Background and Motivation}
Accurate localization and reliable communication are two fundamental aspects of mobile wireless sensor networks \cite{LJTWC}. Localization typically involves determining the spatial coordinates of mobile sensors, while communication facilitates the exchange of information between sensor nodes, enabling data transmission, remote control, and collaboration.  However, the current design in sensor networks often treats localization and communication as separate entities, which can result in high hardware costs and low spectral efficiency (SE). 

Prompted by advanced technologies of the millimeter-wave (mmWave) and massive multiple-input multiple-output (MIMO) in the upcoming 6G, integrated localization and communication systems are driving innovation and research holds in many spectrum and cost limited scenarios \cite{LFJSAC}.  However, obstacles on the ground may prevent the establishment of line-of-sight (LoS) links, leading to severe attenuation of mmWave signals and indicating poor performance in localization and communication. Obtaining reliable connected links of infrastructures is crucial  to support high-quality localization and communication services. Consequently, in areas lacking infrastructures, such as deep canyon environment monitoring, disaster life detection, and border target tracking, unpredictable difficulties may arise in localization and communication \cite{ICC2022}.

Multiple fascinating advantages of unmanned aerial vehicles (UAVs), including high mobility, low cost, and fast deployment, have attracted substantial attention from both academia and industry \cite{WQQJSAC}. UAVs can provide higher reliable LoS connected links and flexible topologies, presenting a cost-effective alternative to traditional infrastructures. Additionally, UAVs have been proven to be significantly important in supporting localization and communication \cite{LYPIOTJ} \cite{LTHTWC}. Nevertheless, the high mobility also inevitably introduces non-negligible Doppler effects, indicating a strong time variant of the channel.

The bi-static channel estimation for communication in multi-antenna systems can be regarded as a standard integrated localization and communication process, where localization parameters, such as distance, time of arrival (TOA) \cite{SYTCOM}, and angle of arrival (AOA) \cite{LTHIOTJ}, can be obtained simultaneously from the channel. However, the increased pilot overhead  will lead to a lower SE, especially in UAV-involved networks.  
Therefore, investigating the frame structure design in UAV-aided localization and communication system  to achieve a high system SE is a particularly challenging task.

\subsection{Related Work}
Many efforts have been dedicated to optimizing the time frame structure for accurate channel estimation and data transmission. The authors in \cite{2021TVT} investigated the channel estimation accuracy with respect to pilot length and base stations density. In \cite{CL2020}, the authors optimized the channel estimation time to improve data rates, where the duration of channel estimation was determined by comparing the predicted and the current data rate. Furthermore, they extended this investigation to the multi-user scenario in \cite{ICC2022}. However, these investigations were limited to static communication conditions.   For dynamic scenarios, the authors in \cite{ICC2017} investigated frame structures to improve the achievable rate in both block and continuous fading, revealing the relationship of channel estimation error and frame structure.  Building on this study, the authors in \cite{CJTWC} proposed a time frame structure optimization problem to maximize the system throughput while guaranteeing the latency, and provided two frame structures for different scenarios. However, these studies focused exclusively on simple single-antenna systems. To address this limitation, the authors in \cite{IOTJ2021} enhanced the sum SE and maximum error probability in the MIMO system by establishing a function between channel coefficients, precoders, error probabilities, and achievable SE. Nevertheless, this study assumed  prefect channel state information (CSI). 
Despite the frame structure and resource allocation strategies designed in the above studies from different perspectives, localization operations are rarely discussed, and three-dimensional (3D) scenarios are not investigated.

Based on the fact that the antenna phased shifts can be expressed as a function of receiver location \cite{CL2022},  
 the authors in \cite{2017TVT} proposed a position-aided channel estimation method for high speed railways,  where the antenna arrays was divided into two part for localization and communication, respectively. They also derived the relationship between mobility and throughput. However, this work required large scale antennas arrays, making it unsuitable for the low-cost wireless sensor networks. In \cite{JSTSP2021}, the authors optimized the pilot overhead by the cooperative beamforming and power allocation. However, this framework was designed for scenarios with sufficient infrastructures. Additionally, the authors in \cite{JSAC20202} utilized a full-duplex UAV as the aerial relay to increase the communication capacity under a 3D channel model. The UAV location was optimized together with the CSI-based beamforming and power control. However, the UAV was assumed to be fixed in the air. 

In practice, the impacts of mobility in UAV-involved network cannot be ignored. In such scenarios, CSI-based beamforming becomes computationally intensive and inaccurate due to the rapidly evolving channel conditions. To handle the channel uncertainty, it will incur a substantial overhead on channel estimation and localization. Location-aware beamforming can be adopted to reduce the beam swapping time and enhance the accuracy of beam alignment.  The authors in \cite{GC2017} investigated a location-aided beam alignment for fast link establishment and provided a robust scheme by Bayesian team decision, but they only considered this framework in a static scenario.    
In \cite{SJ2022}, the authors proposed a location-aware beamforming method for UAV communications by utilizing a spatial estimation method. However, they also assumed perfect CSI.

\subsection{Main Contributions}
Motivated by the aforementioned issues, we aim to reveal the relationship among channel estimation, localization, and communication in UAV aided localization and communication systems. Additionally, we seek to develop an optimization framework based on the findings. The contributions of this paper are summarized as follows. 

\begin{itemize}
  \item We first elaborate on the settings and process of the UAV-aided localization and communication framework. Then, we derive the error bounds for channel estimation and location predication in terms of communication signal-to-noise ratio (SNR) and time frame structure. Moreover, we formulate a joint frame structure, beamforming, and power optimization problem, to maximize the average SE. 
  \item Building on the comprehensive analysis presented, we propose an efficient iterative method to solve the proposed optimization problem. This method employs relaxation and approximation strategies for determining transmission and pilot duration. Remarkably, the beamwidth and power allocation are expressed in closed-form.  
  \item Lastly, we conduct numerical simulations to demonstrate the advantages of our proposed approach compared to four benchmarks. The results showcase that our integrated localization and communication method performs close to the upper bound and achieves over 70\% performance gain compared to the communication-only method.
\end{itemize}

The rest of this paper is organized as follows. Section II introduces the detailed system model and formulates the problem of the UAV-aided localization and communication network. Section III presents an efficient iterative algorithm to solve this optimization problem. Section IV demonstrates the simulation results. Finally, Section V provides the conclusion.

\section{System Model}
\begin{figure}[h]
\centering
\includegraphics[width=1\columnwidth]{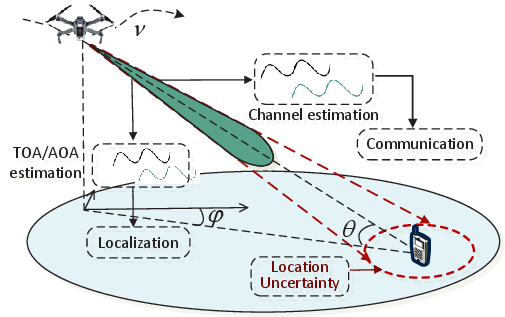}
\caption{UAV aided localization and communication in dynamic scenario.}
\label{scenario}
\end{figure}
We consider a point-to-point UAV-aided localization and communication process depicted in Fig. \ref{scenario}. This involves a UAV equipped with uniform planar array (UPA) employing $N_\text{t}=M\times N$ antennas to assist the localization and communication of the ground node (GN), which is equipped with a single antenna. Without loss of generality, we consider downlink communication in the three-dimensional Cartesian coordinate system. The location and velocity of the UAV at time $t$ are denoted by ${\bf{x}}_\text{u}(t)=\left[p_\text{u}^x(t),p_\text{u}^y(t),p_\text{u}^z(t)\right]^\text{T}$ and ${\boldsymbol{\nu}}_\text{u}(t)=\left[\nu_\text{u}^x(t),\nu_\text{u}^y(t),\nu_\text{u}^z(t)\right]^\text{T}$, respectively.  Similarly, let ${\bf{x}}_\text{n}(t)=\left[p_\text{n}^x(t),p_\text{n}^y(t),p_\text{n}^z(t)\right]^\text{T}$ and ${\boldsymbol{\nu}}_\text{n}(t)=\left[\nu_\text{n}^x(t),\nu_\text{n}^y(t),\nu_\text{n}^z(t)\right]^\text{T}$ denote the location and velocity of the GN, respectively. 

\subsection{Communication Model}
In our considered system, the UAV transmits signal ${\bf{s}}(t)={\bf w}(t)s(t)$ to the GN with power $P(t)$, where ${\bf w}(t)\in \mathbb{C}^{N_\text{t}\times 1}$ is the transmitted beamforming vector, and $\mathbb{E}(|{\bf{s}}(t)|^2)=1$. Thus, the received signal of this multiple-input single-output (MISO) system during the $t$-th time slot can be expressed by  
\begin{equation}\label{rm}
{{r}}(t)=\sqrt{P(t)}{\bf {H}}(t){\bf{s}}(t)+{n}(t),
\end{equation}
where ${\bf H}(t)\in \mathbb{C}^{1\times N_\text{t}}$ is the channel vector between the UAV and the GN, and ${n}(t)$ is the white Gaussian noise at the receiver having zero mean and power $\sigma_0^2$.
The current channel vector can be expressed by 
\begin{equation}\label{rm}
{\bf {H}}(t)=\sqrt{\beta(t)}h(t){\bf{a}}_\text{t}^\text{H}\left(\theta(t),\varphi(t)\right),
\end{equation}
where $\beta(t)=\beta_0/d(t)^2$ is the large scale channel fading with the distance between UAV and GN $d(t)=\sqrt{{\left(p_\text{u}^x(t)-p_\text{n}^x(t)\right)^2+\left(p_\text{u}^y(t)-p_\text{n}^y(t)\right)^2+\left(p_\text{u}^z(t)-p_\text{n}^z(t)\right)^2}}$. The term $\beta_0$ represents the channel power at the reference distance 1 m.  $h(t)$ is small scale Rician fading channel gain\footnote{The channel fading between GN and each antenna element at the UAV are highly correlated. Therefore, we use $h(t)$ to describe the LoS channel.}, and ${\bf{a}}_\text{t}\left(\theta(t),\varphi(t)\right)\in \mathbb{C}^{N_\text{t}\times 1}$ is the array steering vectors of azimuth angle of departure (AoD) $\varphi$ and elevation AoD  $\theta$. 

In this paper, we assume the UAV is equipped with a directional antenna with adjustable beamwidth, which can be expressed by a simplified antenna gain model as \cite{CL2018}
\begin{align}\label{Gt}
\!\!\!\!\!\!G\left(\theta(t), \varphi(t)\right)\!\!=\!\!\left\{
\begin{array}{cc}
\frac{G_0}{\Phi(t)^2},\!\!\!&\varphi-\Phi(t)\!<\!\varphi<\!\varphi+\Phi(t)  \\ & \text{and}\quad \!\!\!\!\theta-\Phi(t)\!<\!\theta<\!\theta+\Phi(t) 
\\0,&\text{otherwise}
\end{array}
\right.,
\end{align}
where $2\Phi(t)\in(0,\pi)$ is the same half-power beamwidth of the antenna on azimuth and elevation, and $G_0$ is a gain constant. The antenna in the GN is omnidirectional with unit gain. Then, the instantaneous SNR can be computed by 
\begin{equation}\label{snr}
\gamma(t)=\frac{P(t)G_0\beta_0}{\Phi(t)^2d(t)^2\sigma_0^2}.
\end{equation}

\begin{figure}[h]
\centering
\includegraphics[width=1\columnwidth]{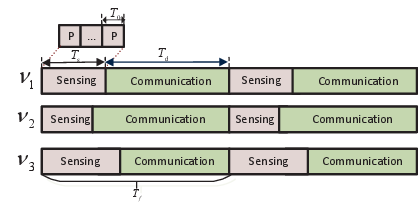}
\caption{Time frame structure of GN for uplink transmission.}
\label{frame}
\end{figure}
The time frame structure is illustrated in Fig. \ref{frame}, where the UAV takes $T_\text{s}$ to estimate the CSI during the sensing phase, which comprises  $k$ orthogonal pilots with the minimum time slot $T_0$. Subsequently, the UAV employs beamforming to build up a communication link with the GN for data transmission. The term $T_\text{d}$ represents the duration of data transmission. The channel is assumed to be correlated in a frame within channel coherence time duration $T_f$, which is influenced by the Doppler effect of the connection link.

\subsection{Channel Estimation}
Due to the time-varying characteristics of the proposed scenario caused by the mobility of the UAV, the CSI will soon become outdated, resulting in the high estimation error.  Typically, the channel estimation error is caused by two factors \cite{TCOM2018}. The fist is the noise, and the second is the rapid temporal variation of channel, reflected in the Doppler effect. 

We denote ${\bf h}_{1:t}=\left[h(1),\cdots,h(t)\right]^\text{T}$ as the true channel vector for time 1 to time $t$, and $\widehat{\bf h}_{1:t}$ as our estimated channel. Then, the channel estimation  error can be expressed as 
\begin{align}\label{CE1}
\Delta{\bf h}_{1:t}=\widehat{\bf h}_{1:t}-{\bf h}_{1:t}=\Delta {\bf h}_{1:t}^\text{n}+\Delta{\bf h}_{1:t}^\text{D},
\end{align}
where  $\Delta {\bf h}_{1:t}^\text{n}$ is the channel mismatch caused by noise, and $\Delta{\bf h}_{1:t}^\text{D}$ is the error due to temporal variation. 

{\bf {\it Proposition 1 :}} Given the pilot length $T_\text{s}$ with $k$ orthogonal symbols, the channel estimation minimum mean square error (MMSE) at time $T_f$, which corresponds to the  $(k+n)$-th symbol, can be expressed as \cite{CJTWC} 
\begin{equation}\label{ce}
\delta_\textbf{h}^2=\underbrace{\frac{1}{1+k\gamma}}_\text{noise}+\underbrace{\frac{2k\gamma\left(1-\alpha_{k,k+n}\right)}{1+k\gamma}}_\text{Doppler},
\end{equation}
where  $\alpha_{k,k+n}$ is the correlation function between the channel at symbol $k$ and symbol $k+n$ with $\alpha_{k,n}=\kappa^{\frac{nf_d}{0.423B}}$. The term $\kappa\in(0,1)$ is the level of correlation,  $f_d$ and $B$ are the Doppler frequency shift and system bandwidth, respectively. 

{\bf {\it Proof :}} See Appendix A
\subsection{Localization Estimation and Prediction}
The motion model of UAV can be expressed as
\begin{align}\label{MM}
{\bf{x}}_\text{m}(t)={\bf{x}}_\text{m}(t-1)+\boldsymbol{\nu}_\text{m}(t)T_0+\boldsymbol\omega(t),
\end{align}
where $\text{m}=\{\text{u},\text{n}\}$ is the index of UAV or GN, and  $\boldsymbol\omega(t)=[\omega_x(t),\omega_y(t),\omega_z(t)]^\text{T}$ is the measurement noise vector along the x-axis, y-axis and z-axis direction, which are assumed to be independent Gaussian distribution with zero mean and variance $\sigma_x^2$, $\sigma_y^2$ and $\sigma_z^2$.

According to our settings, the estimation parameters for localization at time $t$ include time delay $\tau(t)$,  elevation angle $\theta(t)$ and azimuth angle $\varphi(t)$. Therefore, the localization parameter space can be defined as
\begin{align}
\boldsymbol\Theta_t=\left[\tau(t), \theta(t), \varphi(t)\right]^\text{T}.
\end{align}
Let $\widehat{\bf x}_\text{m}$ denote the estimated location of the UAV and GN. In this paper, we use the localization MSE $L_e^2=\mathbb{E}\left\{ \| {\bf x}_\text{m}(t)-\widehat{\bf x}_\text{m}(t)\|^2\right\}$ to evaluate the localization performance, which is determined by the on-broad inertial measurements and the signal measurements between the UAV and GN.

We derive the Posterior Cramér-Rao bound (PCRB) as the lower bound of the estimator variance in dynamic scenarios \cite{ICCC2019}. This metric is valid for a large number of measurements. The equivalent Fisher information matrix (EFIM) of $\boldsymbol\Theta_t$ can be expressed as a block diagonal matrix by chain rules from time 1 to time $t$ with{\footnote{We assume prediction only related to adjacent time slots, indicating ${\bf C}_{i-2,i}=0$. }}
\begin{align}
\mathbf{J}_{\text{m}}^{e}=\!\!\!
\left[\!\!\!
\begin{array}{ccccc}
\mathbf{J}_{1,1}&\mathbf{C}_{1,2}&0&\cdots&0\\
\mathbf{C}_{1,2}^\text{T}&\mathbf{J}_{2,2}&\mathbf{C}_{2,3}&\cdots&0\\
\vdots&\ddots&\ddots&\ddots&\vdots\\
0&\cdots&\mathbf{C}_{t-2,t-1}^\text{T}&\mathbf{J}_{t-1,t-1}&\mathbf{C}_{t-1,t}\\
0&\cdots&0&\mathbf{C}_{t-1,t}^\text{T}&\mathbf{J}_{t,t}\\
\end{array}
\!\!\!\right].
\end{align}
The element of $\mathbf{J}_{i,i}, i\in[1,t]$ is calculated by the FIM $\mathbf{J}_{i,i}^\text{p}$ of measurement pilot signals or information $\mathbf{J}_{i,i}^\text{I}$ from the on-broad inertial measurement unit. Then, we can obtain the expression as follows
\begin{align}\label{ri}
\mathbf{J}_{i,i}=\left\{
\begin{array}{cc}
 \mathbf{J}_{i,i}^\text{p}+\mathbf{J}_{i,i}^\text{I}, & i\leq T_\text{s}, \\
  \mathbf{J}_{i,i}^\text{I}, & T_\text{s}<i,
\end{array}
\right.
\end{align}
where the 3D measurement matric can be characterized by \cite{LTHIOTJ}
\begin{align}
\mathbf{J}_{i,i}^\text{p}=\lambda_{i}^r\mathbf{q}_{i,r}\mathbf{q}_{i,r}^\text{T}+\lambda_{i}^{\theta}\mathbf{q}_{i,\theta}\mathbf{q}_{ij,\theta}^\text{T}+\lambda_{i}^{\varphi}\mathbf{q}_{i,\varphi}\mathbf{q}_{i,\varphi}^\text{T}.
\end{align}
The terms $\lambda_{i}^r$, $\lambda_{i}^\theta$ and $\lambda_{i}^\varphi$ are the ranging information of intensity and angle information from the signal, and $\mathbf{q}_{i,r}$, $\mathbf{q}_{i,\theta}$ and $\mathbf{q}_{i,r}$ are the related  angular vector. The ranging information can be obtained by
$\lambda_{i,r}=\frac{8\pi^2\zeta^2(1-\chi^2)}{c^2}\gamma(i)$, $\lambda_{i,\theta}=\frac{8\pi^2(\zeta\chi+f_c)^2}{c^2}\gamma(i)\sum\limits_{n=1}^{N_t}A_{n,i}$, and $\lambda_{i,\varphi}=\frac{8\pi^2(\zeta\chi+f_c)^2}{c^2}\gamma(i)\sum\limits_{n=1}^{N_t}A_{n,i}$, where $\zeta$, $\chi$, $f_c$ are the related signal parameters, $A_{n,i}$ is the parameter related with the geometry of antenna array \cite{CL2019}.

{\it Proposition 2:} The direction vectors of distance, elevation and azimuth are given by 
\begin{align}
\mathbf{q}_{i,r}=\left[{\begin{array}{*{20}{c}}
{\cos\theta({i})\cos\varphi({i})}&{\sin\theta({i})\cos\varphi({i})}&\sin\varphi({i})
\end{array}}
\right]^\text{T},
\end{align}
\begin{align}
{{\bf{q}}_{i,\theta} }&=\left[\begin{array}{*{20}{c}}
{-\frac{\cos {\varphi({i})}\sin {\theta({i})}}{d({i})}}&{-\frac{\sin {\theta({i})}\sin {\varphi({i})}}{d({i})}}&{\frac{\cos {\theta({i})}}{d({i})}}
\end{array} \right]^\text{T},
\end{align}
\begin{align}
{{\bf{q}}_{i,\varphi}}&=\left[{\begin{array}{*{20}{c}}
{ - \frac{{\sin {\varphi({i})}}}{{{d({i})}\cos {\theta ({i})}}}}&{\frac{{\cos {\varphi({i})}}}{{{d({i})}\cos {\theta ({i})}}}}&0
\end{array}}\right]^\text{T}.
\end{align}

{\it Proof 2:} See Appendix B

The second term in (\ref{ri}) can be expressed as ${\mathbf{J}_{i,i}^\text{I}}={\rm{diag}}\left\{\frac{1}{\sigma_x^2}, \frac{1}{\sigma_y^2}, \frac{1}{\sigma_z^2}\right\}$, and the cross information matrix $\mathbf{C}_{i-1,i}$ is expressed as
\begin{align}
\mathbf{C}_{i-1,i}=-{\rm{diag}}\left\{\frac{1}{\sigma_{x}^2}, \frac{1}{\sigma_{y}^2}, \frac{1}{\sigma_{z}^2}\right\}.
\end{align}
According to the matrix chain and Schur complement rules, we can obtain the localization FIM at time $t$ by 
\begin{align}\label{speb}
{\bf J}({{\bf x}_{\text{m}}(t)})=\!\mathbf{J}_{t,t}\!-\!\mathbf{C}_{t-1,t}({\bf J}({{\bf x}_{\text{m}}(t-1)})\!+\!\mathbf{C}_{t-1,t-1})^{-1}\mathbf{C}_{t-1,t}^\text{T},
\end{align}
in which ${\bf J}({{\bf x}_{\text{m}}(1)})=\mathbf{J}_{1,1}$. Consequently, the localization error can be bounded by 
\begin{align}\label{CRB1}
\mathbb{E}\left\{ \| {\bf x}_\text{m}(t)-\widehat{\bf x}_\text{m}(t)\|^2\right\}\geq \text{tr}\left\{ \left[{\bf J}^{-1}({{\bf x}_{\text{m}}(t)}) \right]\right\}.
\end{align}
 
\subsection{Problem Formulation}
The careful design of frame structure, beamwidth and power allocation has significant impacts on the localization accuracy and channel estimation error, which will contribute to the performance of SE. Given the frame structure and resource allocation, the ergodic achievable SE can be expressed as \cite{JSAC2020} 
\begin{align}\label{SE}
\eta_\text{SE}=\left(1-\frac{T_\text{s}}{T_{f}}\right)\log_2\left(1+\gamma_{e}\right),
\end{align} 
where $\gamma_e=\frac{\gamma\left(1-\delta_\textbf{h}^2\right)}{1+\gamma\delta_\textbf{h}^2}$ is the effective SNR under current instantaneous SNR $\gamma$ with channel estimation error $\delta_\textbf{h}^2$. 

In this paper, we aim to maximize the average SE by optimizing the pilot and data transmission durations, antenna beamwidth, and transmitted power. Denote ${\bf T}_\text{s}=\left[T_\text{s}(1), T_\text{s}(2), \cdots ,T_\text{s}(N) \right]^\text{T}$, ${\bf T}_\text{d}=\left[T_\text{d}(1), T_\text{d}(2), \cdots ,T_\text{d}(N) \right]^\text{T}$, $\boldsymbol\Phi=\left[\Phi(1), \Phi(2), \cdots ,\Phi(N) \right]^\text{T}$
and ${\bf P}=\left[P(1), P(2), \cdots ,P(N) \right]^\text{T}$ as the variable vector of $N$ frames of pilot length, transmission length, beamwidth and power allocation, respectively. 
Then, this optimization problem can be formulated as 
\begin{align}\label{problem}
\underset{{\bf T_\text{s}},{\bf T_\text{d}},\boldsymbol\Phi,{\bf P}}{\max}&\quad \frac{1}{N}\sum_{i=1}^{N}\eta_\text{SE}(i) \\ \tag{\ref{problem}{a}}
 \text{s.t.}\quad  &\text{C1}: \gamma_e(i)\geq \gamma_\text{th} ,\forall i\\ \tag{\ref{problem}{b}}
 &\text{C2}: T_\text{s}(i)>0, T_\text{d}(i)>0 , \forall i\\ \tag{\ref{problem}{c}}
 &\text{C3}: \frac{T_\text{s}(i)}{T_0},\frac{T_\text{d}(i)}{T_0} \in \mathbb{Z}, \forall i\\\tag{\ref{problem}{d}}
 &\text{C4}: \Phi_\text{min}\leq\Phi(i)\leq\Phi_\text{max}, \forall i\\\tag{\ref{problem}{e}}
 &\text{C5}: P(i)>0, \sum_{i=1}^{N}P(i)\leq P_\text{max}, \forall i
\end{align}
where C1 is the communication transmission constraint, C2 and C3 are constraints on pilot and transmission durations, and C4 shows the constraint on beamwidth, limited by the array size. The constraint C5 ensures that the transmitted powers are nonnegative and cannot exceed a maximum value, where $P_\text{max}$ is the total transmitted powers of the UAV for $N$ frames.
Due to the non-convex objective function and the presence of mixed discrete and continuous constraints, this optimization problem can not be optimally solved using current optimization tools.

\section{Efficient Iterative Algorithm}
Since the frame structure, beamwidth and power are highly coupled, obtaining a global optimal solution is challenging. In the section, we aim to find a sub-optimal solution using an efficient iterative algorithm by optimizing one variable while keeping others fixed. Specifically, we first relax the constraints and approximate the objective function to optimize transmission duration. Subsequently, the pilot length is optimized by a specifical solver. Then, the closed-form expression of beamwidth is given using the geometric theory. Next, the optimal power allocation is provided using the Karush-Kuhn-Tucker (KKT) conditions. Finally, we analyze the complexity of overall algorithm.

\subsection{Transmission Duration Determination}
In this subsection, we will determine the transmission duration with fixed other variables. Obviously, the system average SE is maximum when each  frame $i$ is maximized. Consequently, we will analyze the optimization method in one frame, and we omit the index $i$ for simplicity.

Given the pilot length $T_\text{s}$, beamwdith $\Phi$ and power allocation $P$, the $\gamma$ is fixed and the SE of current frame is calculated by 
\begin{equation}\label{P1}
\eta_\text{SE}=\frac{T_\text{d}}{T_\text{s}+T_\text{d}}\log_2\left(\frac{A_1}{A_2-2k\gamma^2\kappa^{T_\text{d}A_3}}\right),
\end{equation}
where $A_1=1+k\gamma+\gamma+k\gamma^2$, $A_2=1+k\gamma+\gamma+2k\gamma^2$ and $A_3=\frac{f_d}{0.423BT_0}$. 

To tackle the non-convexity of (\ref{P1}), we employ the successive convex optimization method to approximate it with a convex form using the first-order Taylor expansion around a given local point $T_\text{d}^r$. Therefore, the approximated objective function can be expressed  as
\begin{equation}
\tilde{\eta}_\text{SE}^r=A_4^r\left(T_\text{d}-T_\text{d}^r\right)+A_5^r,
\end{equation}
where 
\begin{align}
A_4^r=&-\frac{T_\text{d}^r}{T_\text{s}+T_\text{d}^r}\log_2e\left(\frac{2k\gamma^2\ln\kappa A_3\kappa^{T_\text{d}^rA_3}}{A_2-2k\gamma^2\kappa^{T_\text{d}^rA_3}}\right)\\
&+\frac{T_\text{s}}{T_\text{s}+T_\text{d}^r}\log_2\left(\frac{A_1}{A_2-2k\gamma^2\kappa^{T_\text{d}^rA_3}}\right),
\end{align}

\begin{equation}
A_5^r=\frac{T_\text{d}^r}{T_\text{s}+T_\text{d}^r}\log_2\left(\frac{A_1}{A_2-2k\gamma^2\kappa^{T_\text{d}^rA_3}}\right).
\end{equation}

Moreover, there exists a maximum transmission duration $T_\text{d}^\text{max}$, which is calculated by C1 as follows
 \begin{align}\label{maxtd}
  \gamma_\text{th} \leq \frac{\gamma\left(1-\delta_\textbf{h}^2\right)}{1+\gamma\delta_\textbf{h}^2}, \quad
   \delta_\textbf{h}^2 \leq \frac{\gamma-\gamma_\text{th}}{\gamma+\gamma\gamma_\text{th}}.
 \end{align}

Plugging (\ref{ce}) into the equation (\ref{maxtd}), we can obtain the maximum transmission duration by 
 \begin{equation}
 T_\text{d}^\text{max}= \frac{\ln\left(\frac{2k\gamma+1}{1+k\gamma}-\frac{\gamma-\gamma_\text{th}}{\gamma+\gamma\gamma_\text{th}}\right)A_4}{\ln(\kappa) f_d},
 \end{equation}
where $A_4=\frac{1+k\gamma}{2k\gamma}T_0$.  

The constraint C3 renders the integer limitations of $T_\text{d}^r$. To address this, we relax this constraint into continuous space. Consequently, the optimization problem for one frame structure can be replaced by the following problem 
\begin{align}\label{problem1}
\underset{{T_\text{d}}}{\max}&\quad \tilde{\eta}_\text{SE}^r \\ 
 \text{s.t.}\quad  & 0<T_\text{d}\leq T_\text{d}^\text{max}. \nonumber
\end{align}
In this way, the optimization problem becomes a standard convex optimization problem, enabling us to solve it via CVX solver. Then, we can obtain the transmission duration $T_\text{d}^*=T_0\lfloor\frac{T_\text{d}}{T_0}\rfloor$

\subsection{Pilot Length Optimization}
Similar to determining transmission duration, the system's average SE is maximized by ensuring each frame's SE reaches its peak. Therefore, we will also optimize the pilot length in one frame while keeping other variables fixed. The average SE can be expressed as
\begin{equation}\label{P2}
\eta_\text{SE}=\frac{n}{k+n}\log_2\left(\frac{k\gamma^2+k\gamma+B_1}{2k\gamma^2B_2+k\gamma+B_1}\right),
\end{equation}
where $n$, $k$ are the symbol numbers of transmission and pilot, respectively. The term $B_1=\gamma+1$ and $B_2=1-\alpha_{n,k+n}$. Maximizing $\eta_\text{SE}$ is equivalent to minimizing $-\eta_\text{SE}$. Then this subproblem can be reformulated as
\begin{align}\label{problem2}
\underset{T_\text{s}}{\min}&\quad -{\eta}_\text{SE} \\ 
 \text{s.t.}\quad  & T_\text{s}>0, \nonumber\\
 & k=\frac{T_\text{s}}{T_0}, k \in \mathbb{Z}. \nonumber
\end{align}

Due to the complicated objective function (\ref{P2}), this is a standard nonlinear integer programming problem (NLIP), which is difficult to solve directly. 
In this paper, we address this NLIP problem using NOMAD solver in MATLAB OPTI Toolbox, which can find solutions in several iterations. After that, we can obtain the optimal pilot length $T_\text{s}^*$.

\subsection{Beamwidth Design}
When the pilot and transmission duration are fixed, the channel estimation error decreases by increasing the SNR. According to our antenna model in (\ref{Gt}), a narrower beamwidth corresponds to higher antenna gain. However, if the $\theta$ and $\varphi$ are out of the beam, the antenna gain trends toward zero. That indicates that high localization accuracy offers the high beam alignment and small beamwidth, which contributes the high performance on the SNR of the connected link. Similar to the pervious descriptions, the average SE is maximized when the SE of each frame is maximized. Here, we provide a {\it Proposition} to demonstrate the relationship between beamwidth and localization accuracy.

{\it Proposition 3 \cite{LTHTWC}:} Given data transmission duration $T_\text{d}$, pilot length $T_\text{s}$ and transmitted power $P$, we can obtain the localization accuracy of UAV and GN by (\ref{CRB1}), denoted by $L_\text{u}^2$ and $L_\text{n}^2$, respectively. Denote horizontal distance between UAV and GN as $d_h=\sqrt{{\left(p_\text{u}^x(t)-p_\text{n}^x(t)\right)^2+\left(p_\text{u}^y(t)-p_\text{n}^y(t)\right)^2}}$. Then, the optimal beamwidth can be calculated in the following two cases.

1. Case 1: When $d_{h}\geq l_\text{u}+2l_\text{n}$
\begin{align}\label{BW1}
\Phi^*=&\max\left\{\arctan\left(\frac{d_1}{d_z}\right)-\arctan\left(\frac{d_2}{d_z}\right), \Theta_\text{min} \right\},
\end{align}
where $d_1={d_h-l_\text{u}+2l_\text{n}}$, $d_2=d_h-l_\text{u}-2l_\text{n}$ and $d_z=p_\text{u}^z-p_\text{n}^z$. 

2. Case 2: When $d_{h}\leq l_\text{u}+2l_\text{n}$
\begin{align}\label{BW2}
\Phi^*=&\max\left\{2\arctan\left(\frac{2l_\text{n}}{p_\text{u}^z-p_\text{n}^z}\right), \Theta_\text{min} \right\}.
\end{align}

{\it Proof:} The similar proof illustration can be found in the Appendix B of our pervious investigation \cite{LTHTWC}. However, unlike the proof in our previous research, we consider the imperfection in UAV location in this paper. Therefore, the optimal beamwidth should introduce the uncertainty of UAV location.

\subsection{Power Allocation}
Given other variables, a higher transmitted power will lead to a higher SNR, better localization accuracy, and a lower channel estimation error. Thus, the average SE monotonically increases with the transmitted power. Therefore, the maximum SE is achieved when the constraint C5 becomes an equality. To ensure the constraint, we can obtain the lower bound of transmitted power $P_\text{min}(i)$ at each frame $i$ by solving following equation.
\begin{equation}
  D_1\gamma(i)^2+\gamma_\text{th}\left(k+1\right)\gamma(i)+\gamma_\text{th}=0,
\end{equation}
where $D_1=2kB_2\gamma_{th}+2kB_2-k$. This equation can be easily proven to have one positive root. Then, this non-linear convex subproblem can be formulated as an equivalent form 
\begin{align}\label{problem4}
\underset{\bf P}{\min}&\quad -\sum_{i=1}^{N}\eta_\text{SE}(i) \\ 
 \text{s.t.}\quad  & P_\text{min}(i)-P(i)\leq 0, \forall i \nonumber \\ 
 &\sum_{i=1}^{N}P(i)- P_\text{max}=0, \forall i \nonumber
\end{align} 
Under this formulation, the power allocation of each frame can be determined using Lagrangian multiplier method with KKT conditions. The Lagrangian function for this problem is expressed as
\begin{align}
\mathcal{L}({\bf P}, {\boldsymbol \lambda}, \mu)=-\sum_{i=1}^{N}\eta_\text{SE}(i)+&\sum_{i=1}^{N}\lambda(i)\left( P_\text{min}(i)-P(i)\right) \nonumber
\\&+\mu\left(\sum_{i=1}^{N}P(i)- P_\text{max}\right),
\end{align}
where $\lambda(i)\geq0$ and $\mu$ are the lagrangian multipliers. According to the KKT conditions, we have 
 \begin{align}\label{KKT}
 \left\{
\begin{aligned}
 &-\eta_\text{SE}'(i)-\lambda(i)+\mu=0, \quad  \forall i\\
& \lambda(i)\left( P_\text{min}(i)-P(i)\right)=0, \quad  \forall i\\
& \sum_{i=1}^{N}P(i)- P_\text{max}=0,\\
\end{aligned}
\right.
 \end{align}
where $\eta_\text{SE}'(i)=\frac{\partial\eta_\text{SE}(i)}{\partial P(i)}$ is the first derivative of $\eta_\text{SE}(i)$.
To guarantee the solvability of this problem, the maximum power should satisfy $P_\text{max}\geq NP_\text{min}$.

In the second equation of (\ref{KKT}),  if $P_\text{min}(i)\neq P(i)$, then $\lambda(i)=0$. By substituting $\lambda(i)$ into the first equation and combining it with the equations of the first condition in (\ref{KKT}) and the third condition, we can solve for the $i+1$ variables with $i+1$ equations. Then, we can obtain the power allocation result of each frame with $P(i)^\dag$. Consequently, the optimal power  of each frame $i$ is given by 
\begin{equation}\label{OP}
  P(i)^*=\max\left\{P(i)^\dag, P_\text{min}\right\}.
\end{equation}

\subsection{Overall Algorithm and Complexity Analysis}
The overall iterative algorithm is summarized in {\bf Algorithm 1}. Here, we present the analysis of the complexity for our proposed method. Denote $I$ as the number of iteration for convergency. For optimizing the transmission duration, the major complexity is solving $N$ approximated convex problems with complexity $N\mathcal{O}(c(1))$, where $C(1)$ represents the computational complexity by employing interior-point method with 1 variable \cite{convex}. For optimizing the pilot length, the major complexity lies in solving the NLIP with complexity $N\mathcal{O}(\log_2k(i))$, where $k(i)$ is the pilot length of frame $i$.  The beamwidth and power allocation are determined by the closed-form. Consequently, the total complexity of our proposed algorithm is $N\mathcal{O}(c(1)+\log_2k(i))$.
\begin{algorithm}
\caption{Proposed Efficient Iterative Algorithm}
\label{alg1}
\begin{algorithmic}[1]
\STATE {Set the initial state$\left({\bf T}_\text{d}^0, {\bf T}_\text{s}^0, \boldsymbol\Phi^0, {\bf P}^0 \right)$ and the iteration number $I=1$.}
\STATE {\bf Repeat}
\FOR {$i=1:N$}
\STATE{Given $\left({\bf T}_\text{s}^{I-1}, \boldsymbol\Phi^{I-1}, {\bf P}^{I-1}\right)$, solve problem (\ref{problem1}) for frame $i$ to obtain the optimal transmission duration
 $T_\text{d}^*$, update ${\bf T}_\text{d}^{I}(i)=T_\text{d}^*$.}
 \ENDFOR
\FOR {$i=1:N$}
\STATE{Given $\left({\bf T}_\text{d}^{I}, \boldsymbol\Phi^{I-1}, {\bf P}^{I-1}\right)$, solve problem (\ref{problem2}) for frame $i$ to obtain the optimal pilot length
 $T_\text{s}^*$, update ${\bf T}_\text{s}^{I}(i)=T_\text{s}^*$.}
 \ENDFOR
 \FOR {$i=1:N$}
\STATE{Given $\left({\bf T}_\text{d}^{I}, {\bf T}_\text{s}^{I}, {\bf P}^{I-1}\right)$,  obtain the optimal beamwidth $\Phi^*(i)$ of frame $i$ by (\ref{BW1}) (\ref{BW2}), update $\boldsymbol\Phi^{I}(i)=\Phi^*(i)$.}
 \ENDFOR
\STATE{Given $\left({\bf T}_\text{d}^{I}, {\bf T}_\text{s}^{I}, \boldsymbol\Phi^{I}\right)$, obtain the optimal power allocation result by (\ref{OP}), update into ${\bf P}^{I}$.}
\STATE {{\bf {Until}} the objective function $\sum_{i=1}^{N}\eta_\text{SE}(i)$ converges.}
\end{algorithmic}
\end{algorithm}

\section{Numerical Simulations}

In this section, we thoroughly evaluate the proposed UAV aided localization and communication method via numerical simulations. The size of the simulation scenario is $1000m$$\times$$200m$$\times$$100m$. Unless otherwise specified, the main simulation parameters are presented in Table \ref{table1}.

\begin{table}[h]
\centering
\caption{Main Simulation Parameters}\label{table1}
\begin{tabular}{|c|c|}
  \hline
  Parameters & Value \\ \hline
  Symbol period &  $T_0=66.7$ [$\mu \text s$] \\
  Carrier frequency &$f_c=4.9 $ [GHz]\\
  Effective bandwith & $\zeta=1$ [MHz]\\
  Antenna reference gain&$G_0=2.28$ \cite{CL2018}\\
  Baseband carrier correlation & $\chi=\sqrt{0.32}$\cite{ZTTTWC}\\
  Noise power & $\sigma_0^2=-110$ [dBm] \cite{LDCL} \\
  The channel gain at 1 $\rm{m}$ & $\beta_0=-80$ [dB]\\
  Threshold of SNR & $\gamma_\text{th}=3$ [dB] \\
  Number of antenna& $N_\text{a}=8$\\
  Minimum beamwidth & $\Phi_\text{min}=5^{\circ}$\\
  Motion noise &$\sigma_x=\sigma_y=\sigma_z=0.005$ $\rm{m}$\\
  Channel correlation level & $\kappa=0.8$\\
  \hline
\end{tabular}
\end{table}

To highlight the advantages of our framework, we compare the proposed efficient iterative method with a lower bound and three benchmark schemes in the following.
\begin{itemize}
  \item {\bf Upper bound:} In this scheme, we employs an exhaustive method for this joint optimization problem, where the beamwidth and power allocation are executed using closed-form expressions in (\ref{BW1}) (\ref{BW2}) and (\ref{OP}).
  \item {\bf Benchmark 1:} In this scheme, we optimize the transmission duration, pilot length and power allocation without considering the localization. This optimization problem is also solved using the iterative algorithm. 
  \item {\bf Benchmark 2:} This scheme combines localization operation with a fixed frame structure. The beamwidth and power allocation are also given by our derivations.
  \item {\bf Benchmark 3:} In this scheme, we adopt the heuristic algorithm of particle swarm optimization (PSO) to find
 suboptimal solutions for the optimization method. The related parameters of this algorithm can be found in \cite{LTHICC}.
\end{itemize}

\begin{figure}[h]
\centering
\includegraphics[width=1\columnwidth]{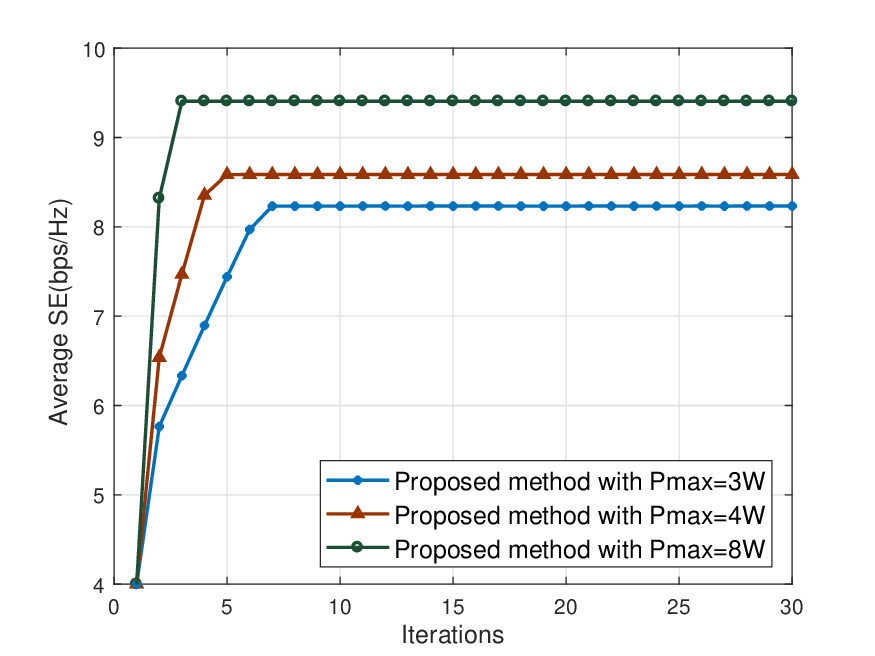}
\caption{Convergence performance of proposed method for various maximum total transmitted power.}
\label{convergency}
\end{figure}

We first analyze the convergence behavior of our proposed iterative algorithm for various maximum transmitted power settings. This analysis is conducted with $N=100$ and a UAV-GN relative velocity of $50 \rm{m/s}$. As illustrated in Fig. \ref{convergency}, the average SE of GN always converges within 10 iterations. Higher transmitted power levels, indicating the higher SNR, lead to faster convergence rates. Specifically, under the high SNR condition with $P_\text{max}=8\rm{W}$, our method converges within 3 iterations.

We also conducted a comparative evaluation of the computational time between the proposed method and PSO-based method with 200 particles. Using the same simulation platform, the proposed method only takes about 5 seconds, while the PSO-based method and the exhaustive research require over 50 and 300 seconds, respectively. This demonstrates the notable computational efficiency of our method. 

\begin{figure}[h]
\centering
\includegraphics[width=1\columnwidth]{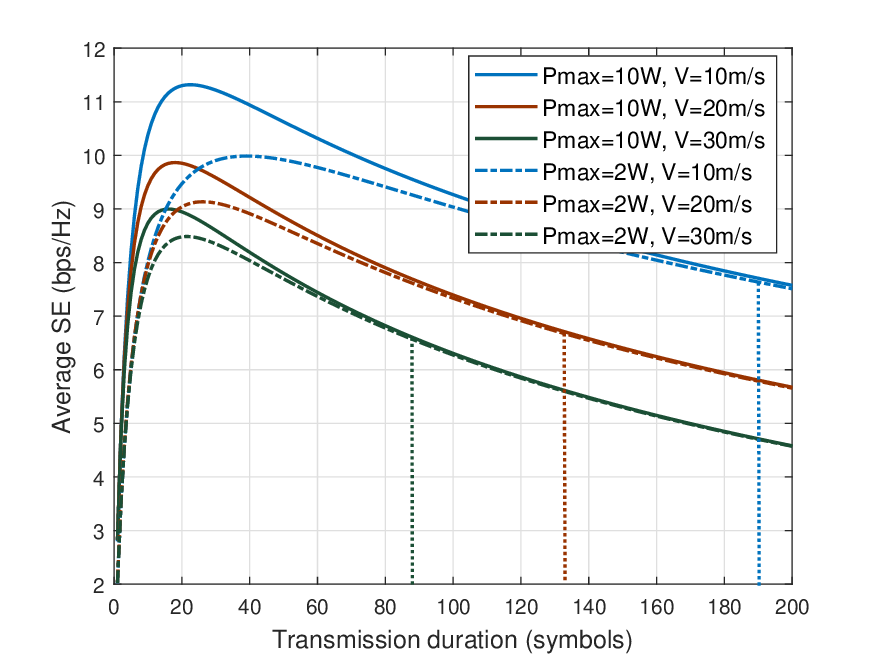}
\caption{Average SE performance of different conditions with fixed pilot length versus transmission duration.}
\label{SEPV}
\end{figure}
\begin{figure}[h]
\centering
\includegraphics[width=1\columnwidth]{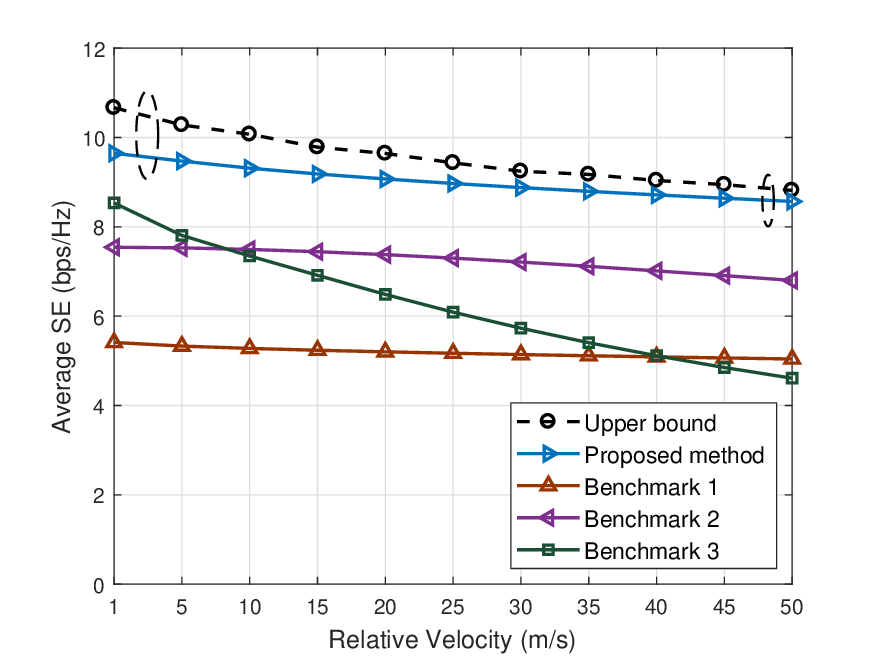}
\caption{Average SE performance of different methods versus UAV-GN relative velocities.}
\label{SEvsV}
\end{figure}
\begin{figure}[h]
\centering
\includegraphics[width=1\columnwidth]{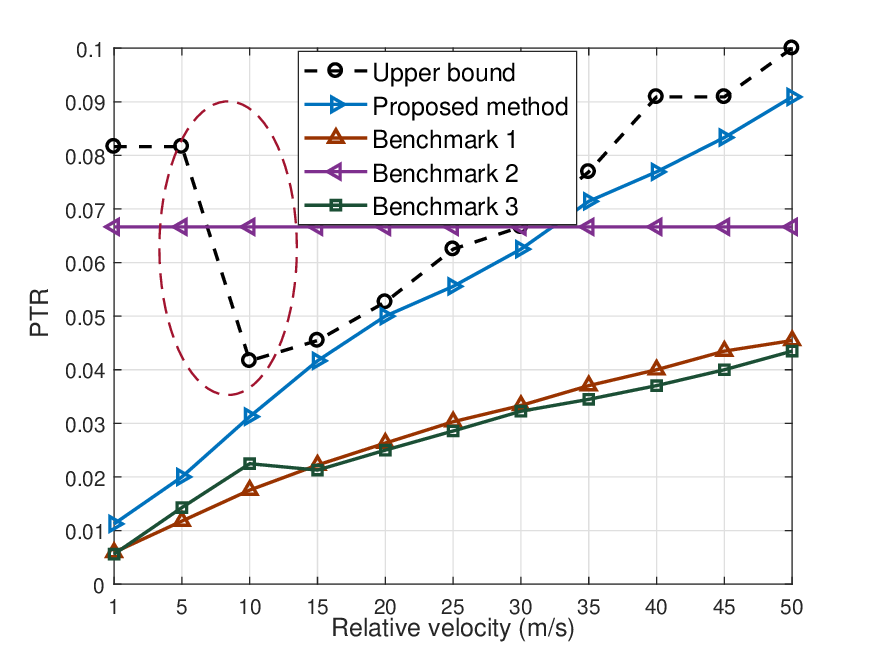}
\caption{PTR of different methods versus UAV-GN relative velocities.}
\label{PTRvsV}
\end{figure}

Fig. \ref{SEPV} demonstrates the performance of the average SE with a fixed pilot length ($k=5$) under different power limitations and relative velocities. As shown in this figure,  higher power limitations lead to better performance, and there exists an optimal transmission duration $T\text{d}^*$ for each condition.
When $T_\text{d}<T_\text{d}^*$, the average SE increases with transmission duration due to the increase in the first term ($\frac{T_\text{s}}{T_\text{s}+T\text{d}}$) of (\ref{SE}). When $T_\text{d}\geq T_\text{d}$, the average SE decreases with transmission duration due to the worsened channel estimation and location prediction results. An interesting observation is that as the  transmission duration increases, different power limitations converge in the same level. That can be explained by the effective SNR equation $\gamma_e=\frac{\gamma\left(1-\delta_\textbf{h}^2\right)}{1+\gamma\delta_\textbf{h}^2}$, which can be rewritten as $\gamma_e=\frac{\left(1-\delta_\textbf{h}^2\right)}{\frac{1}{\gamma}+\delta_\textbf{h}^2}$. Therefore, the effective SNR is highly dependent on the significant channel estimation error $\sigma_\textbf{h}$.  As observed in this figure (colored straight dotted-line), higher maximum power provides a greater range for convergence.

In Fig. \ref{SEvsV}, we evaluate the average SE performance across different UAV-GN relative velocities, comparing it against the upper bound and other benchmarks. Here, we set the maximum transmitted power $P_\text{max}=4\rm{W}$.  The results demonstrate that our proposed method perform close to the upper bound, meaning that it is near-optimal, and achieves a remarkable over 75\% performance gain compared to {benchmark 1}, highlighting the advantages of integrated localization and communication, which also aligns with the motivation of our investigation.   Furthermore, our method shows significant performance gain compared to other benchmarks. As velocities increase, our results also display the greater robustness than {benchmark 3}, which fails to achieve reliable performance under lower effective SNR conditions.

Fig. \ref{PTRvsV} shows the pilot and transmission duration ratio (PTR) of different methods with the same conditions in Fig. \ref{SEvsV}. From the upper bound and our proposed method, we can find that our integrated localization and communication system increases the PTR compared to the communication-only method. This improvement is attributed to the fact that more pilot overhead can provide better localization accuracy, leading to a narrower beamwidth and higher SNR.  There exists a change of the upper bound at the low velocity region. The PTR decreases at relative velocity $10\rm{m/s}$. This is due to at this situation the SE is mainly influenced by the first term $\frac{T_\text{s}}{T_\text{s}+T\text{d}}$ of (\ref{SE}). That also verifies the performance gap in Fig. \ref{SEvsV} between our method and the upper bound.
As the velocity increasing, the channel estimation error becomes grievous. The PTR should increase to improve the performance localization and channel estimation.  The PTR of Our method perform close to the upper bound at the high Doppler effect regions. 
In comparison to the fixed frame method, our proposed method can reduce the pilot overhead to guarantee high SE when the Doppler effect is small, while increasing the pilot ratio to compensate for the channel estimation mismatch when the Doppler effect is rigorous.

\begin{figure}[h]
\centering
\includegraphics[width=1\columnwidth]{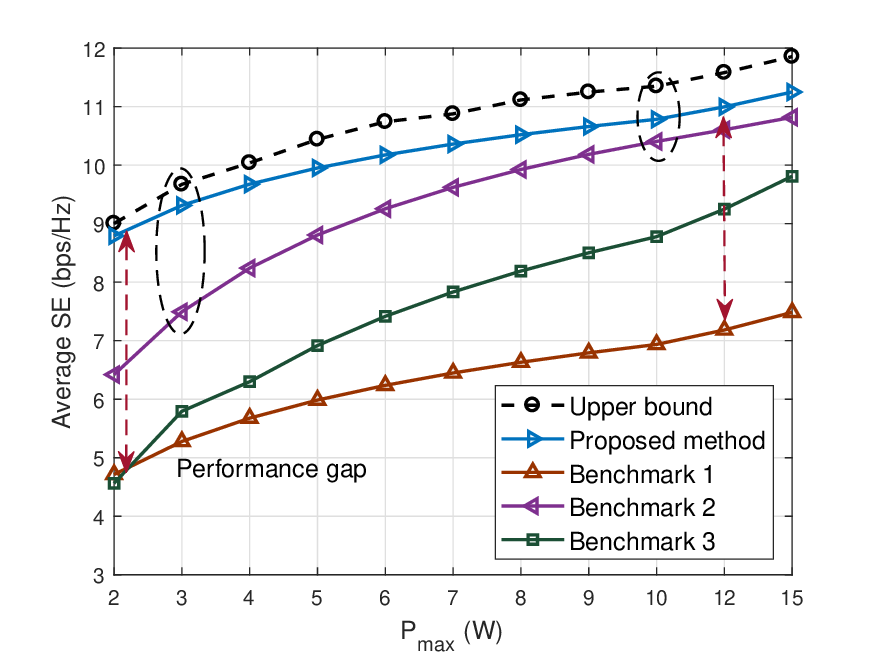}
\caption{Average SE performance of different methods versus maximum transmitted powers.}
\label{SEvsP}
\end{figure}
\begin{figure}[h]
\centering
\includegraphics[width=1\columnwidth]{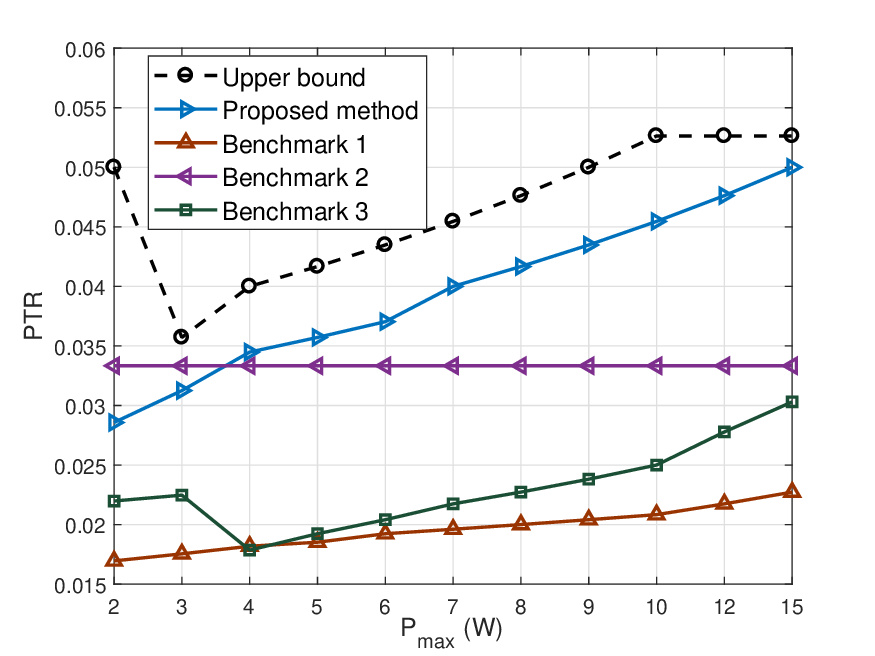}
\caption{PTR of different methods versus maximum transmitted powers.}
\label{PTRvsP}
\end{figure}

In Fig. \ref{SEvsP}, we evaluate the average SE of different methods with various transmitted power limitations. Here, we set the relative velocity at $10\rm{m/s}$, and the PTR of fixed frame is adjusted to $1/30$ in the benchmark 2. As expected, the average SE exhibits a positive correlation with the total transmitted power, and our proposed method consistently maintains a small gap with the upper bound. The benchmark 2 also demonstrates the good system performance, highlighting the importance of localization operation. As can be observed, the performance gap between our method and other benchmarks is smaller with the increasing SNR. That phenomenon showcases the effectiveness of our proposed method in low SNR regions. 

In Fig. \ref{PTRvsP}, we analyze the PTR for the simulation results in Fig. \ref{SEvsP}. Different from the result in Fig. \ref{PTRvsV}, where increasing the pilot overhead mitigates the Doppler effect, this result reveals that it also needs to add the pilot ratio or reduce the duration of transmission as transmitted power increases. A shorter data transmission time implies reduced channel time-variant characteristics, indicating that the channel estimation interval has a significant impact compared to SNR in channel estimation. At high SNR levels, the time frame structure tends to become fixed.

\section{Conclusion}
In this paper, we explored a UAV-aided localization and communication system. We formulated a joint frame structure, beamwidth, and power allocation optimization problem, to maximize the average SE of UAV-GN link. 
Given this  non-convex problem, we developed an efficient iterative algorithm to obtain sub-optimal solutions. Base on the numerical results, we can draw following key conclusions (i) The proposed method closely approaches the upper bound with remarkably low complexity, outperforming the communication-only method by over 70\%  in average SE. (ii) The Doppler effect significantly influences channel estimation error, and the PTR should increase in both high dynamic and high SNR scenarios. (iii) For a specifical scenario, there exists a unique frame structure to achieve optimal average SE. These findings provide valuable insights for system design. In the future, the multi-user, computing analysis, and trajectory management can be investigated.

\section*{Appendix A: Proof of Proposition 1}
The channel estimation is executed by the receiving pilot symbols. Denote ${\bf x}_k=\left[x(1),x(2),\cdots,x(k)\right]^\text{T}$ and ${\bf y}_k=\left[y(1),y(2),\cdots,y(k)\right]^\text{T}$ be the input training and received symbol vectors, respectively. According to the MMSE rule \cite{ICC2017}, the estimated channel can be expressed as 
\begin{equation}
\widehat{h}_k=\sqrt{\gamma}(1+k\gamma)^{-1}{\bf x}_k^\text{H}{\bf y}_k,
\end{equation}
where $(\cdot)^\text{H}$ is the Hermitian operation. According to the received signal model in (1), $y(k)=\sqrt{\gamma}{{h}}(k){x}(k)+{n}(k)$. In this paper, we assume  $x(k)$ has unit mean square, which means $|{\bf x}_k|^2=k$. Then, the estimated channel can be rewritten
\begin{equation}
\widehat{h}_k=\frac{k\gamma h(k)}{1+k\gamma}+\frac{\sqrt{\gamma}{\bf x}_k^\text{H}{\bf n}_k}{1+k\gamma},
\end{equation}
where ${\bf n}_k=\left[n(1),n(2),\cdots,n(k)\right]^\text{T}$ is the noise vector. 

In this paper, the temporal channel correlation between two symbols can be modeled as a first-order Markov process \cite{CJTWC}
\begin{equation}
{h}_n=\prod_{i=k+1}^{n}\alpha_{k,i}h_k+\sqrt{1-\prod_{i=k+1}^{n}\alpha_{k,n}^2}\varsigma_{n},
\end{equation}
where $\varsigma_{n}$ is a Guassian random variable with zero mean and unit variance, and $\alpha_{k,n}=\kappa^{\frac{f_d}{0.423B}}$ is the correlation factor between $h_k$ and $h_n$. The term $\kappa$ is the level of correlation,  $f_d$ and $B$ are the Doppler frequency shift and system bandwidth, respectively.

Therefore, given $k$ pilot symbols, the channel estimation error at the symbol $n$ can be calculated by
\begin{align}\label{error}
\delta_{\bf h}&={h}_n-\widehat{h}_k=\prod_{i=k+1}^{n}\alpha_{k,i}h_k \nonumber\\ 
&+\sqrt{1-\prod_{i=k+1}^{n}\alpha_{k,n}^2}\eta_{n}-\frac{k\gamma h(k)}{1+k\gamma}+\frac{\sqrt{\gamma}{\bf x}_k{\bf n}_k}{1+k\gamma}.
\end{align}

We take the MSE of (\ref{error}) to obtain $\mathbb{E}\left\{|\delta_{\bf h}|^2\right\}$ with
\begin{align}
\delta_{\bf h}^2&=\left(\prod_{i=k+1}^{n}\alpha_{k,i}-\frac{k\gamma}{1+k\gamma}\right)\mathbb{E}\left\{|h_k|^2\right\} 
\nonumber\\
&+\left({1-\prod_{i=k+1}^{n}\alpha_{k,i}^2}\mathbb{E}\left\{|\eta_{n}|^2\right\}\right)+\frac{{k\gamma}}{\left(1+k\gamma\right)^2}\nonumber
\\ &=\frac{1}{1+k\gamma}+{\frac{2k\gamma\!\!\left(1-\!\!\prod\limits_{i=k+1}\limits^{n}\!\!\alpha_{k,i}\right)}{\left(1+k\gamma\right)}}\nonumber\\
&=\frac{1}{1+k\gamma}+{\frac{2k\gamma\!\!\left(1-\alpha_{k,k+n}\right)}{\left(1+k\gamma\right)}}.
\end{align}

Thus the proof of {\it Proposition 1} is completed. 

\section*{Appendix B: Proof of Proposition 2}
We convert the parameter space ${\boldsymbol\Theta}_t$ to the position coordinates, and show their relationships as follows.
\begin{equation}
\left\{ {\begin{array}{*{20}{l}}
c\tau_t = \sqrt{\left(p_\text{u}^x(t)-p_\text{n}^x\right)^2+\left(p_\text{u}^y(t)-p_\text{n}^y\right)^2+\left(p_\text{u}^z(t)-p_\text{n}^z\right)^2}\\
\varphi_t = \arctan\left(\frac{p_\text{u}^y(t)-p_\text{n}^y(t)}{p_\text{u}^x(t)-p_\text{n}^x(t)}\right)\\
\theta_t = \arctan\left(\frac{p_\text{u}^z(t)-p_\text{n}^z(t)}{\sqrt{\left(p_\text{u}^x(t)-p_\text{n}^x\right)^2+\left(p_\text{u}^y(t)-p_\text{n}^y\right)^2}} \right)
\end{array}} \right..
\end{equation}

We omit $t$ for simplicity. Without loss generality, we derive the direction vector for localization of GN as follows.
\begin{equation}
\renewcommand{\arraystretch}{2}
\setlength{\arraycolsep}{1.2pt}
  {\bf q}_{r}=\left[\begin{array}{c}
                        \frac{\partial\tau}{p_\text{n}^x} \\
                        \frac{\partial\tau}{p_\text{n}^y} \\
                        \frac{\partial\tau}{p_\text{n}^z} 
                      \end{array}
  \right]=\left[\begin{array}{c}
                       - \frac{p_\text{u}^x-p_\text{n}^x}{d} \\
                        -\frac{p_\text{u}^y-p_\text{n}^y}{d} \\
                        -\frac{p_\text{u}^z-p_\text{n}^z}{d} 
                      \end{array}
  \right]=\left[\begin{array}{c}
                       - \cos\varphi\cos\theta \\
                        -\sin\varphi\cos\theta \\
                        -\sin\theta 
                      \end{array}
  \right],
\end{equation}
\begin{equation}
\renewcommand{\arraystretch}{2}
\setlength{\arraycolsep}{1.2pt}
  \!\!\!\!\!\!\!\!{\bf q}_{\theta}=\left[\begin{array}{c}
                        \frac{\partial\theta}{p_\text{n}^x} \\
                        \frac{\partial\theta}{p_\text{n}^y} \\
                        \frac{\partial\theta}{p_\text{n}^z} 
                      \end{array}
  \right]\!\!=\!\!\left[\begin{array}{c}
                       \frac{\left(p_\text{u}^x-p_\text{n}^x\right)\left(p_\text{u}^z-p_\text{n}^z\right)}{\sqrt{\left(p_\text{u}^x-p_\text{n}^x\right)^2+\left(p_\text{u}^y-p_\text{n}^y\right)^2}d} \\
                        \frac{\left(p_\text{u}^y-p_\text{n}^y\right)\left(p_\text{u}^z-p_\text{n}^z\right)}{\sqrt{{\left(p_\text{u}^x-p_\text{n}^x\right)^2+\left(p_\text{u}^y-p_\text{n}^y\right)^2}}d} \\
                        -\frac{{\left(p_\text{u}^x-p_\text{n}^x\right)^2+\left(p_\text{u}^y-p_\text{n}^y\right)^2}}{d} 
                      \end{array}
  \right]\!\!=\!\!\left[\begin{array}{c}
                       \frac{\cos\varphi\sin\theta}{d} \\
                        \frac{\sin\varphi\sin\theta}{d} \\
                        -\frac{\cos\theta}{d}
                      \end{array}
  \right],
\end{equation}
\begin{equation}
\renewcommand{\arraystretch}{2}
\setlength{\arraycolsep}{1.2pt}
  \!\!\!\!\!\!\!\!{\bf q}_{\varphi}=\left[\begin{array}{c}
                        \frac{\partial\varphi}{p_\text{n}^x} \\
                        \frac{\partial\varphi}{p_\text{n}^y} \\
                        \frac{\partial\varphi}{p_\text{n}^z} 
                      \end{array}
  \right]\!\!=\!\!\left[\begin{array}{c}
                       \frac{p_\text{u}^y-p_\text{n}^y}{\sqrt{\left(p_\text{u}^x-p_\text{n}^x\right)^2+\left(p_\text{u}^y-p_\text{n}^y\right)^2}} \\
                        -\frac{p_\text{u}^x-p_\text{n}^x}{\sqrt{{\left(p_\text{u}^x-p_\text{n}^x\right)^2+\left(p_\text{u}^y-p_\text{n}^y\right)^2}}} \\
                       0 
                      \end{array}
  \right]\!\!=\!\!\left[\begin{array}{c}
                       \frac{\sin\varphi}{d\cos\theta} \\
                        -\frac{\cos\varphi}{d\cos\theta} \\
                        0
                      \end{array}
  \right].
\end{equation}

 Then, the proof of {\it proposition 2} is completed.
\bibliographystyle{IEEEtran}
\bibliography{ref}

\end{document}